# Broadband Radar Invisibility with Time-Dependent Metasurfaces

Vitali Kozlov, Dmytro Vovchuk, and Pavel Ginzburg

*Abstract*— **Concealing objects from interrogation has been a primary objective since the integration of radars into surveillance systems. Metamaterial-based invisibility cloaking, which was considered a promising solution, did not yet succeed in delivering reliable performance against real radar systems, mainly due to its narrow operational bandwidth. Here we propose an approach, which addresses the issue from a signal-processing standpoint and, as a result, is capable of coping with the vast majority of unclassified radar systems by exploiting vulnerabilities in their design. In particular, we demonstrate complete concealment of a 0.25 square meter moving metal plate from an investigating radar system, operating in a broad frequency range approaching 20% bandwidth around the carrier of 1.5GHz. The key element of the radar countermeasure is a temporally modulated coating. This auxiliary structure is designed to dynamically and controllably adjust the reflected phase of the impinging radar signal, which acquires a user-defined Doppler shift. A special case of interest is imposing a frequency shift that compensates for the real Doppler signatures originating from the motion of the target. In this case the radar will consider the target static, even though it is moving. As a result, the reflected echo will be discarded by the clutter removal filter, which is an inherent part of any modern radar system that is designed to operate in real conditions. This signal-processing loophole allows rendering the target invisible to the radar even though it scatters electromagnetic radiation. This application claims the benefit of patent number 273995, filed on April 16 2020.**

## I. Introduction

The invention of radars was soon followed by extensive research and development of counter measures, the most well-known being the invention of stealth technology [1]. By employing special geometric designs and carefully selected materials, a reduction of the target's radar cross section and the resulting backscattered energy was successfully achieved, substantially reducing distances between the radar and the target for successful detection. In addition to stealth technology, numerous active jamming countermeasures have been developed [2]. In this case, carefully engineered signals are transmitted to the investigating radar systems, which then processes them as if they are real echoes, causing the system to wrongly conclude such estimation parameters as range, velocity and number of targets in the scene. Less sophisticated systems content with blinding the interrogating radar by transmitting random noise, which serves to degrade the signal to nose ratio in the receiver [3], [4]. The field of meta-materials [5]–[11] (to some extend adopting ideas from frequency selective surfaces, previously developed by the radio frequency community), has further advanced the ability to cloak objects of interest from investigation by introducing man-made materials to tailor electromagnetic scattering [12]. Numerous approaches were undertaken [13]–[17], giving rise to the concept of an invisibility cloak, yet all of the passive methods suffered from narrow bandwidth, polarization sensitivity and bulky designs [18], reducing the practical application of such devices. In addition to narrow bandwidth and other operational shortcomings, these cloaks tended to lose their properties when moving with respect to the radar. To overcome this problem, a 'Doppler cloak' was suggested theoretically as a way to restore the invisibility of moving cloaked objects [19]. The implementation of such Doppler cloaks was theorized to be possible in the continuous wave case by using a metasurface with time-dependant scattering properties [20], yet the suggested metasurface was still narrowband. To the best of the author`s knowledge, this manuscript presents the first experiment conducted to demonstrate the Doppler cloak (a micro-Doppler cloak was experimentally shown [21]). The investigation ahead takes a few steps forward, showing theoretically and demonstrating experimentally that practical broadband radar invisibility is achievable with semi-passive (non-radiating) Doppler cloaks by exploiting an often-overlooked vulnerability of most contemporary radars.

Modern radar systems can simultaneously measure the location and radial velocity of investigated targets. In the simplest terms, their method of operation is based on transmitting modulated (for example pulsed) electromagnetic radiation towards a target and recording the reflected echoes [22]–[24]. From the delay between the transmitted and received signals (time of flight) the range to the target can be deduced, while the phase difference between consecutive pulses, produced by the Doppler effect, allows the measurement of the instantaneous radial velocity. Our semi-passive approach to

Vitali Kozlov is with Tel Aviv University, Tel Aviv, 6997801 Israel. (e-mail: vitaliko@mail.tau.ac.il) – *corresponding author*

Dmytro Vovchuk is with Tel Aviv University, Tel Aviv, 6997801 Israel and Yuriy Fedkovych Chernivtsi National University, Chernivtsi, 58012 Ukraine (e-mail: dimavovchuk@gmail.com).

Pavel Ginzburg is with Tel Aviv University, Tel Aviv, 6997801 Israel. (e-mail: pginzburg@tauex.tau.ac.il) .

radar invisibility does not require transmitting signals to confuse or jam the radar, nor does it require a lot of a priori knowledge about the type of radar at hand. Instead, a temporally modulated reflecting coating is suggested, which can control the time dependent phase of the electromagnetic field as it is backscattered towards the interrogating radar. The proposed invisibility concept is depicted on Fig. 1.

Owing to the fact that the Doppler information is extracted from the difference in the phase of consecutive pulses, dynamically controlling the reflected phase from the target produces backscattered echoes, which contain fake Doppler signatures that are indistinguishable from the ones created by genuine motion. It is therefore possible to deceive a radar system into concluding it is observing a moving target when the target is in fact stationary. More importantly, the proposed method makes it possible to compensate the real phase difference between consecutive pulses, which originates from the movement of the target, thereby cloaking the signatures of motion and making the target appear stationary to the radar. It is worth stressing that the Doppler-cloaked target still scatters a lot of energy since it is not employing any method of scattering suppression. The proposed method only serves to deny the interrogating system information about the target's instantaneous velocity, which is crucial for the proper operation of numerous radar systems relying on clutter removal methods, such as the moving target indicator (MTI) filter [25]–[28]. The absence of Doppler information originating from the target will make it indistinguishable from the surrounding clouds, mountains and ground, which can backscatter much more energy with very small Doppler shifts. This means that the MTI filter will remove the energy related to the target along with the rest of the clutter, rendering it effectively invisible to the radar.

For narrowband radars, typically defined as those having less than 5% bandwidth around the carrier frequency, it is sufficient to cloak the Doppler signature around the carrier. For broadband signals, on the other hand, the Doppler shift in the entire range should be cloaked, requiring broadband phase matching as will be discussed ahead.

The manuscript is organized as follows: first, a theoretical outline of the metasurface Doppler cloak is presented and its broadband cloaking capabilities are analyzed. Additional pathways to increasing the operation bandwidth are further discussed before the experimental section. Then an experimental implementation is demonstrated, showing close agreement with the theoretical model.

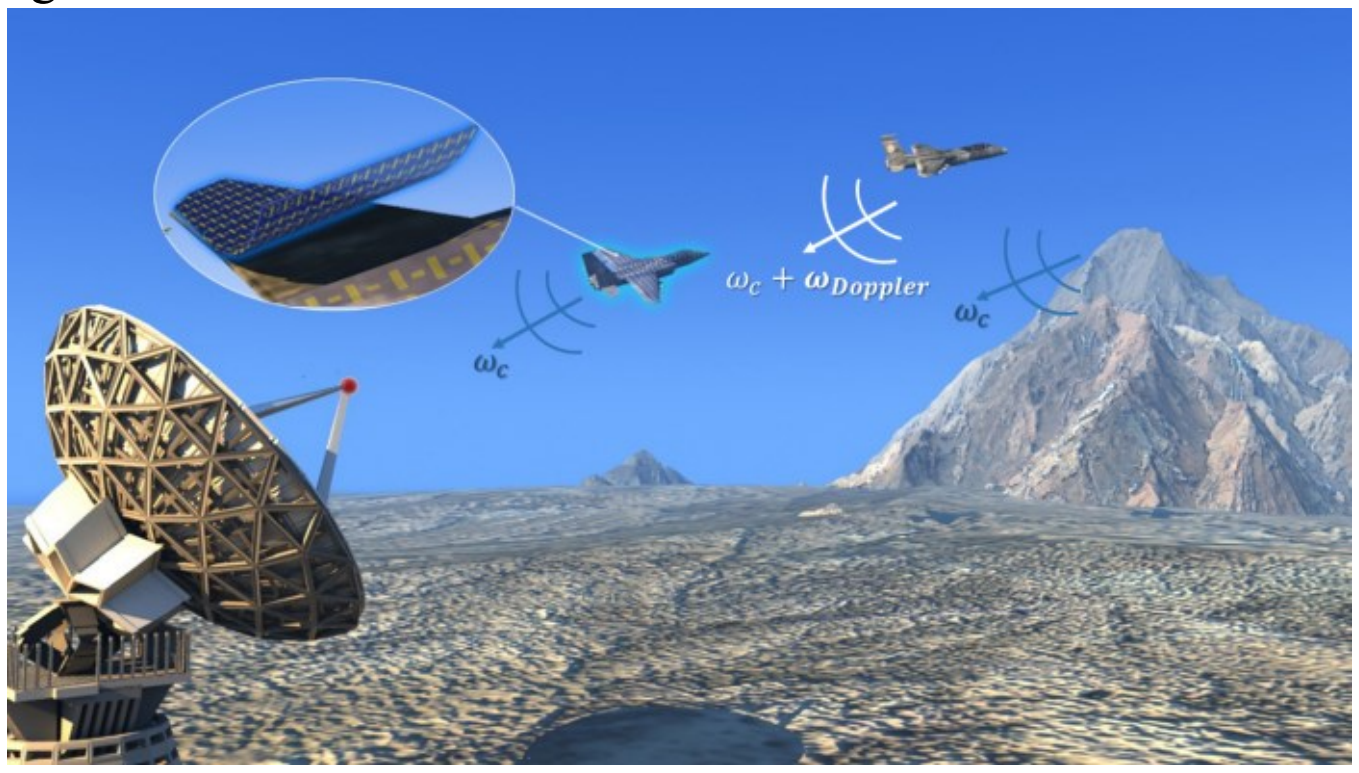

Fig. 1. Illustration of the proposed invisibility concept. Two airplanes and a mountain are shown reflecting echoes back into an interrogating radar. The left plane is coated with a special metasurface, which dynamically controls the phase of the scattered electromagnetic pulse back into the radar. Temporal modulation of the metasurface causes the reflected phase to remain unchanged in time, even though the airplane is moving. This serves to conceal the Doppler signature of the aircraft, making it appear as stationary as the surrounding clutter (i.e. the ground or the mountains near the radar). This causes the radar to filter out the cloaked target together with the stationary clutter, rendering it invisible. The second airplane appears clearly against the backdrop of the mountain, even though it is a much smaller scatterer, due to the fact it can be detected using its Doppler signature.

## II. Theory and Modelling

In order to understand the operation of the metasurface Doppler cloak, it is instructive to consider a single dipole, which allows gaining physical insight into the phenomenon. This insight will be used to understand the basic operation principle behind the suggested broadband invisibility concept.

### A. *Controlling the scattered phase from a single dipole*

It is well known that the polarizability of a dipole has a Lorenzian shape in the frequency domain [29], where near the resonance the phase is approximately linear in frequency. The dipole is excited by an incident radiation, which is partially reflected back into the source (e.g. a radar antenna). If the resonant frequency of the dipole it temporally modulated, the scattered field acquires an additional time-dependent phase shift. Note that radar systems almost never rely on the amplitude of scattered echoes for detection, mainly due to its unpredictability in real environments and unidentified targets. Temporal modulation of the dipole is realized by incorporating a voltage-controlled capacitor (varactor) within the structure. Fig. 2(a) demonstrates a lumped elements scheme for the scattering scenario containing the dipole and varactor. Here the impinging wave is represented by the voltage source $V_{in}$, while the resistance $R$, capacitance $C$, and inductance $L$ characterize the dipole and depend on the material composition and geometric shape of the resonator. Placing a voltage dependant varactor in the feeding gap of the dipole can serve as a resonance-shifting element, allowing control over the scattered phase, shown in Fig. 2(b). The varactor is represented on the scheme as an additional capacitor $C_v$ in parallel to the dipole's natural one ($C$). The current flowing through the resistor is related to the scattered electromagnetic field and its phase is the goal of the following derivation. The current in the frequency domain is given by:

$$i = \frac{j\omega\left(C + C_v(t)\right)V_{in}}{1 + j\omega R\left(C + C_v(t)\right) - \omega^2 L\left(C + C_v(t)\right)} \tag{1}$$

where $\omega$ is the angular frequency of the impinging radiation. It is important to note that a time scale separation method is used to derive (1) [30]. Here the assumption is that any time-dependent changes in the varactor's capacitance are far slower than the carrier frequency of the exciting radiation. In this case, the dipole may be considered as stationary at any particular time, solving the fast scattering problem while keeping the capacitance $C_v$ as a parameter. As it will be shown ahead, the required modulation frequency of the varactor is of no more than a few kHz, while typical radar systems transmit above 1 GHz, making this approximation perfectly justifiable.

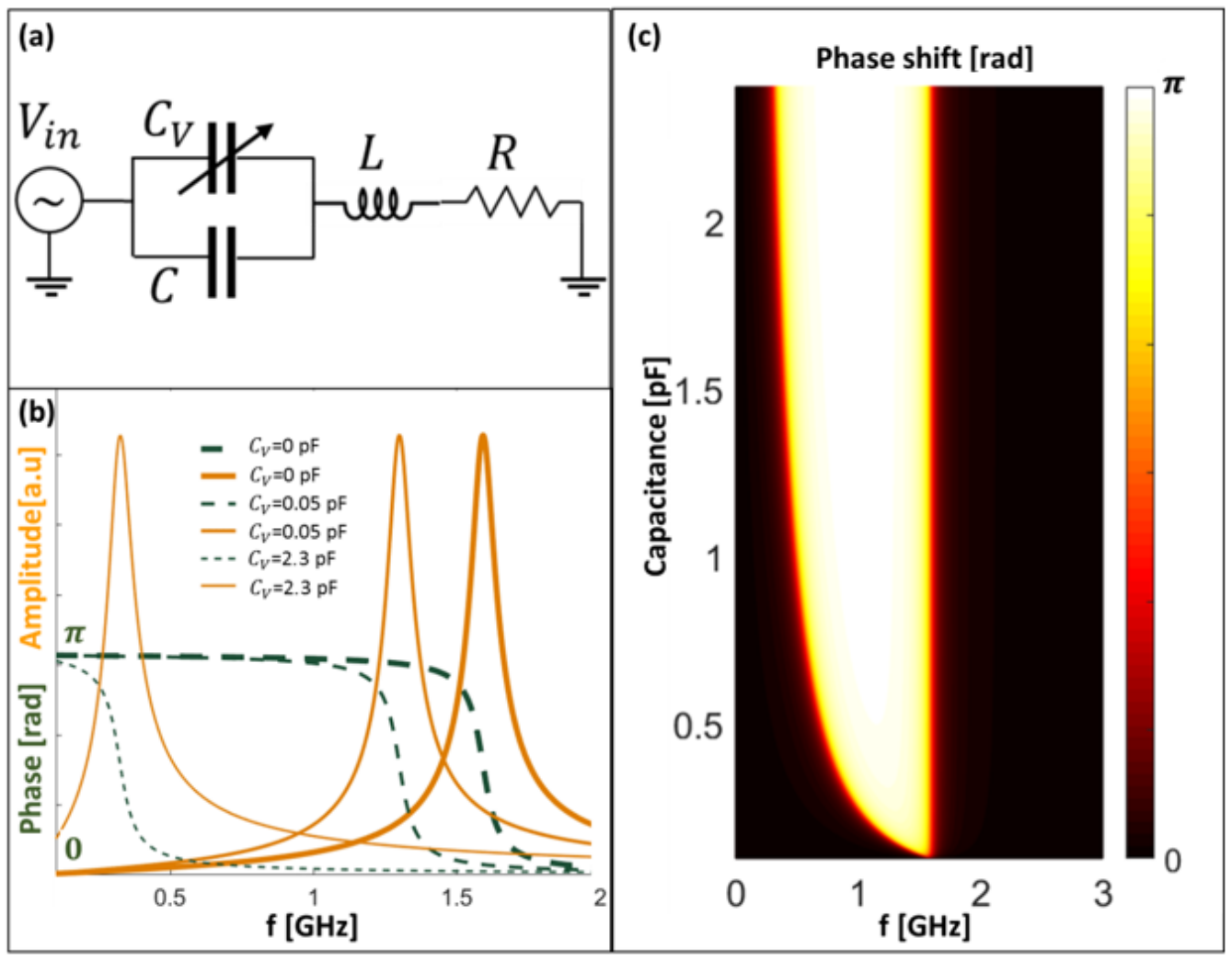

Fig. 2. Controlling the scattered phase from a dipole using a varactor. (a) The equivalent circuit of the scattering scenario - a dipole is represented by internal inductance, Ohmic + radiation resistance $R$, capacitance $C$ and an additional varactor capacitor $C_v$ , which is introduced within the structure. (b) Resonance control - tuning the bias voltage drop on the varactor serves to shift the resonance of the dipole, which affects the amplitude and phase of the scattered field. Orange lines – amplitudes of the current on the wire, green dashed lines – corresponding phases. Nominals of the varactor are given in the legends. (c) Color map of the achievable phase shift of the scattered field as a function of varactor capacitance and incident wave frequency.

The phase of the current in (1) is given by:

$$\varphi = \tan^{-1}\left(\frac{\mathrm{Im}\{i\}}{\mathrm{Re}\{i\}}\right) = \tan^{-1}\left(\frac{\omega^2 L(C+C_v(t))-1}{\omega R(C+C_v(t))}\right) \quad (2)$$

Fig. 2(b) summarizes the results of (2), demonstrating that the phase undergoes rapid change from $\pi$ to 0 around the resonant frequency, which is controlled by the varactor capacitance. Other system parameters used for the plot are: $C$ = 0.1 pF, $L$ = 0.1 μH and $R$ = 50Ω. It is important to note that these values were chosen in order to obtain a response similar to the one found in experiment, yet this combination is obviously not unique. Since radar systems are not sensitive to absolute phase, but rather to the phase difference between consecutive pulses, it is more instructive to inspect the change in phase as a function of varactor capacitance:

$$\Delta\varphi(C_v(t),\omega) = \varphi(t) - \varphi(t) =$$
$$\tan^{-1}\left(\frac{RC_v(t)\omega}{1+\left(R^2C(C+C_v(t)) - L(2C+C_v(t))\right)\omega^2 + L^2C(C+C_v(t))\omega^4}\right) \quad (3)$$

where $C_v(0)$ is assumed to be 0 for simplicity. Plotting (3) versus the incident frequency $f = \omega/(2\pi)$ produces Fig. 2(c), where a knife-like image may be seen. Any horizontal cut-line of the knife represents a phase shift versus frequency, similar to the plots shown on Fig. 2(b) (specifically subtraction between thin and bold dashed green lines). On the other hand, controlling the bias of the voltage drop on the varactor diode while keeping the incident frequency constant, is equivalent to taking a vertical cut-line in Fig. 2(c). It can be seen that with increasing varactor capacitance the phase goes from 0 to $\pi$ abruptly in an almost step-like manner, with the transition capacitance depending on the incident frequency. This transition capacitance is located along the knife's edge, and its accurate conditions will be derived later on. The conclusion from the above discussion is that for any frequency of incident radiation, it is possible to continuously induce up to a $\pi$ phase shift in the reflected field by carefully tuning the bias voltage of the varactor diode. While this $\pi$-shift can severely hamper the ability of an investigating system to deduce instantaneous velocity, full control over $2\pi$ phase is desirable for achieving complete invisibility, as will be discussed ahead.

## B. *Achieving 2π phaseshift*

As demonstrated above, a single dipolar scatterer is not sufficient to provide full phase control over the reflected wave, motivating the development of more advanced configurations. Hereinafter we show that resonator-based reflect arrays, often termed as metasurfaces, indeed can allow controllable $2\pi$ phase shift of the reflected waves. A typical example is a structure with a switchable characteristic impedance, which has properties resembling either perfect electric or a perfect magnetic conductor. In this case the reflection coefficient varies from '-1' to '1' respectively and thereby allowing to obtain full control over the reflected phase [20], [31]. While analytic models for arrays of scattering dipoles do exist [32], [33], they might be quite cumbersome for obtaining immediate physical insights. In addition, these models tend to neglect higher-order multipolar interaction, edge effects in finite sized systems and several other aspects, which might be important in practical realizations. Instead, it is frequently preferable to use full wave numeric simulations in order to optimize the metasurface and obtain the desired results. This is the approach undertaken ahead using the time domain FDTD method implemented in CST Microwave Studio. Fig. 3(a) shows a single unit cell of a 9x9 array of dipoles on a dielectric substrate (FR-4, $\varepsilon_r$ = 4.3), which is located above a metallic surface, assumed to be a perfect electric conductor. The metallic surface represents the target that is to be cloaked. A biasing network, made out of thin wires, provides the temporal modulation of the voltage drop, which is used to control the capacitance of the varactor. An additional capacitor $C_\omega(\omega)$ is shown in parallel to the varactor, however it will only be required later on and is assumed to be disconnected in the following discussion. Fig. 3(b) shows a color map of the phase shift of the reflected field as a function of substrate thickness (distance between the dipole and the metal surface, denoted as '$h$' on Fig. 3(a)) and incident frequency. The map is obtained by repeating the full wave simulation for different substrate thicknesses, while discretely switching the capacitance between the maximum and minimum values of 2.6 pF and 0.6 pF, in accordance with the datasheet of the varactor diode that is used in the experimental set up. A full $2\pi$ phase shift is thus clearly achievable for substrate thicknesses between about 2 and 15 millimeters, when the other dimensions are fixed at $d$ = 50mm, $l$ = 40mm, $l_s$ = 42mm, $w_s$ = 10 mm, $w$ = 10mm, the dipole width is 1mm and the width of the biasing network wires is much smaller than 1mm. It is worth noting that this numerical optimization of the entire structure allows approaching experimental realization quite closely, as will be seen later.

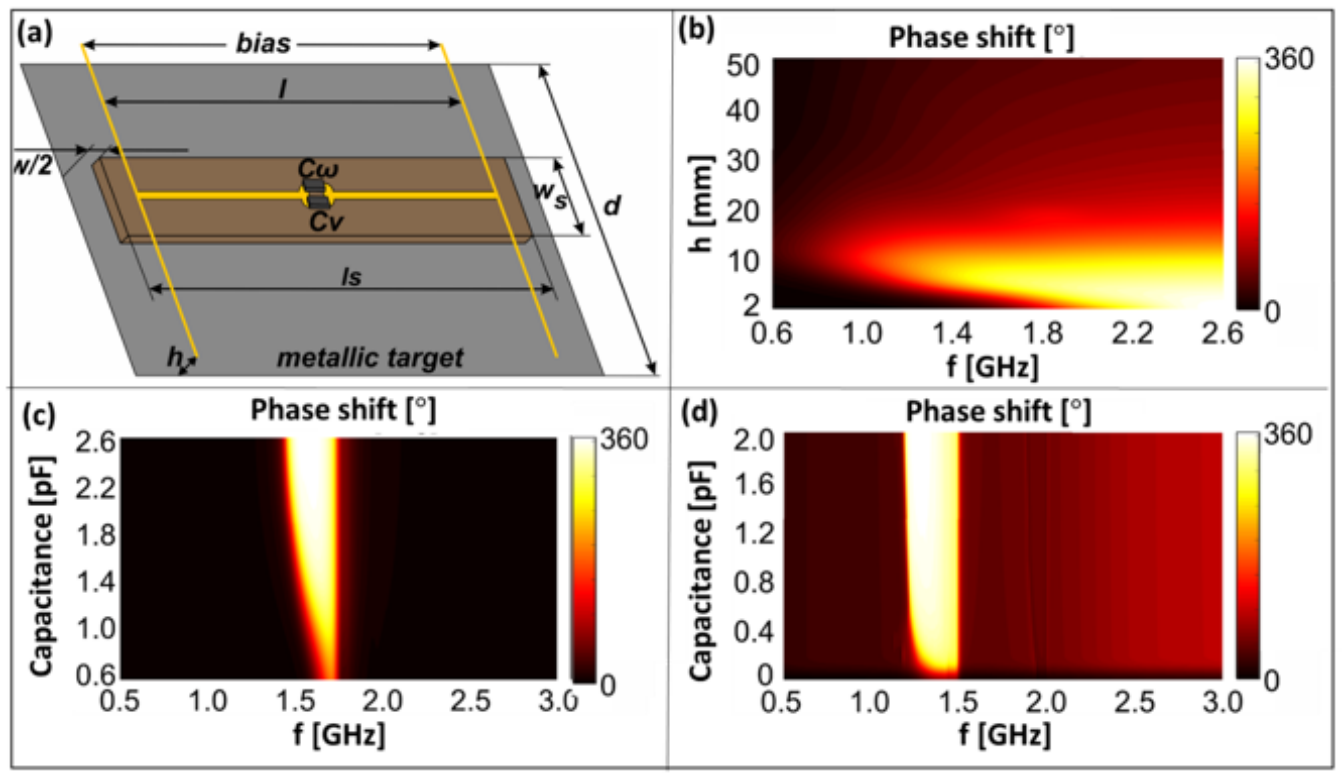

Fig. 3. Numerical modelling of the metasurface Doppler cloak, constructed from a 9x9 array of dipoles on a substrate above a metal ground plane which represents the target to be concealed. (a) A single unit cell containing the dipole, varactor and biasing network. (b) Relative phase shift as a function of distance h between the dipole and metallic target, as well as the frequency of the incident radiation. (c) The phase shift as a function of varactor capacitance and incident frequency for dielectric width $h$ = 6mm (d) Improving the operational bandwidth - the phase shift from the metasurface after adding an additional frequency dependant capacitor $C_\omega$, shown on Fig. 3(a).

Fig. 3(c) shows the phase shift of the metasurface as a function of the incident wave frequency and the varactor's capacitance, bearing a remarkable resemblance to Fig. 2(c). Indeed, a knife-like image can be seen on both plots. The key difference, however, lies in the fact that a maximal controllable phase shift of $2\pi$ can be obtained with this array, unlike the $\pi$ phase shift attainable with a single dipole. The resulting array is similar to the phase switched screen [34], which was previously used to redistribute the reflected energy outside of the receiving radar band, therefore making it less visible to the interrogator. This was achieved by fast switching of reflectivity between two values, causing broadband modulation of the incident field. Our report, however, investigates a perturbative and quasi-static approach, which does not significantly modulate the frequency of the incident wave – this is highly important in the case of wide-band radar systems and provides significantly better performances in passive deception applications, since the low frequency modulation does not radiate at the switching frequency. The purpose of our modulation is to produce a linear time dependent phase shift of the backscattered field, which exactly compensates for the linear phase shift produced by the motion of a target via the Doppler effect. The linearity of the phase can be achieved by modulating the bias voltage in time with the inverse function of the capacitance-phase relation shown on the vertical cuts of Fig. 3(c), which serves to 'straighten out' the phase dependence on time at the frequency of interest. However, linear phase response is not retained across the entire band, seeing as the threshold varactor capacitance, which is the edge of the knife in Fig. 3(c), varies from frequency to frequency. Additional correction must be taken to achieve broad phase matching.

### *C. Broadband phase matching*

In order to achieve broadband performance, it is necessary to control the phase change across a range of frequencies with an identical (shared along the band) driving varactor bias voltage. The time dependant varactor biasing approach, summarized in Fig. 3(c), shows that the discussed metasurface might not cover the entire bandwidth of a wideband radar system (typically defined as above 5% around the carrier). The reason is that for a range of frequencies the phase difference transitions from 0 to $2\pi$ occurs at different varactor capacitance values. In order to achieve a broadband response, the knife-shape of Fig. 3(c) should be transformed into a rectangular form, where the transition from 0 to $2\pi$ occurs at the same cutoff values of the varactor capacitance, leading to broadband phase matching. To accomplish this goal, an additional frequency-dependent capacitor is introduced within the circuit in parallel with the varactor, as shown on Fig. 3(a). The goal of this new element is to 'straighten out the knife' by shifting the frequency dependent threshold capacitance toward lower varactor values. In the case of the single dipole discussed earlier, this transitional cutoff capacitance $C_\omega(\omega)$ may be derived from (3). by updating the lumped elements scheme to include the new dispersive element $C_v \rightarrow C_v + C_\omega$ and setting $\Delta\varphi = \pi/4$ as a convenient but arbitrary value for the threshold, after which the phase difference transitions from 0 to $2\pi$. It is then possible to solve and obtain an expression for $C_\omega$:

$$C_\omega(\omega) = \frac{1+\left(R^2C^2\omega^2 + \left(LC\omega^2 - 2\right)LC\omega^2\right)}{RC\omega - R^2C^2\omega^2 - LC\omega^2\left(LC\omega^2 - 1\right)}C \tag{4}$$

Analysis of (4) shows that $C_\omega$ is a decaying function of frequency (for the values of $L$, $C$ and $R$ that were used before) and its plot versus frequency is in fact the edge of the knife shown on Fig. 2(c). For the array of dipoles, on the other hand, no simple formula exists and a numerical approach must be undertaken in a similar fashion by using the color map of Fig.3(c) in order to obtain the threshold capacitance. Fig. 3(d) shows a simulation, which is identical to the one performed in Fig. 3(c), with the exception of an additional frequency dependent capacitor that was placed in the gap of each dipole in the array. The frequency dependant capacitor was modelled as two subwavelength metallic plates with a frequency dependant dielectric material in between. The material demonstrates anomalous dispersion, which resembles the leading edge of the "knife's edge", and provides, in fact, the required rectifying capacitance $C_\omega$. It can be seen that after adding this capacitor, around 300MHz of bandwidth become available for simultaneous phase switching. This represents more than 20% of bandwidth around the carrier (assuming the carrier is in the middle of the band). An additional shift of the rectified knife in comparison with the knife shape in Fig. 3(c) is observed. This is due to the addition of the new capacitor, and is expected due to the fact the dispersive capacitance values do not vanish anywhere in the observed frequency band, serving to lower the resonant frequency of the array. It is also worth noting that the impedance of the dispersive capacitor does not obey Foster's reactance theorem and $-1/(\omega C_\omega)$ is decreasing with frequency. In the above discussion we proposed using dispersive dielectric materials between the capacitor's electrodes. Furthermore, active circuits are already proven to provide these desired capabilities [35].

## III. Experiment

In order to demonstrate the capabilities of the described metasurface at concealing a large target from an investigating radar, an experimental device was fabricated and is shown on Fig. 4. The array of 9x9 dipoles was manufactured according to the simulation presented in the previous section with the same dimensions. The dipoles were chemically etched from a copper surface that was deposited on top of a dielectric FR-4 substrate. SMV1405 varactors were soldiered in the dipole gaps, while the edges of the dipoles were soldiered to thin wires forming the biasing network. The array of dipoles was glued on top of a supporting structure, which was transparent to centimetre waves and served as a spacer of 15mm, altogether forming the metasurface. This metasurface was designed to be placed in front of the metal plate, which was the target to be hidden from the radar. The target covered by the metasurface was placed on a polyester structure that connected it to a motorized conveyor belt, which enabled moving it forward and backward with a controllable speed reaching up to about 0.04m/s. A stepped frequency continuous wave (SFCW) radar system was implemented with a Network Analyzer, which is capable of sweeping the entire band of interest (1.2-1.7GHz) while recording the amplitude and phase of the received echoes from the target. This type of radar is typically used in ultra-wideband applications since it is able to transmit carriers sequentially, while the receiver is locked on the transmitted frequency in a predefined time window. This architecture allows avoiding expensive high frequency samplers that would otherwise be needed for sampling extremely short pulses [36]. The radar's antenna was placed directly in front of the moving conveyor inside an anechoic chamber and served both for transmitting and receiving the radiation (monostatic radar scheme), linearly polarized in the direction of the dipoles (horizontal).

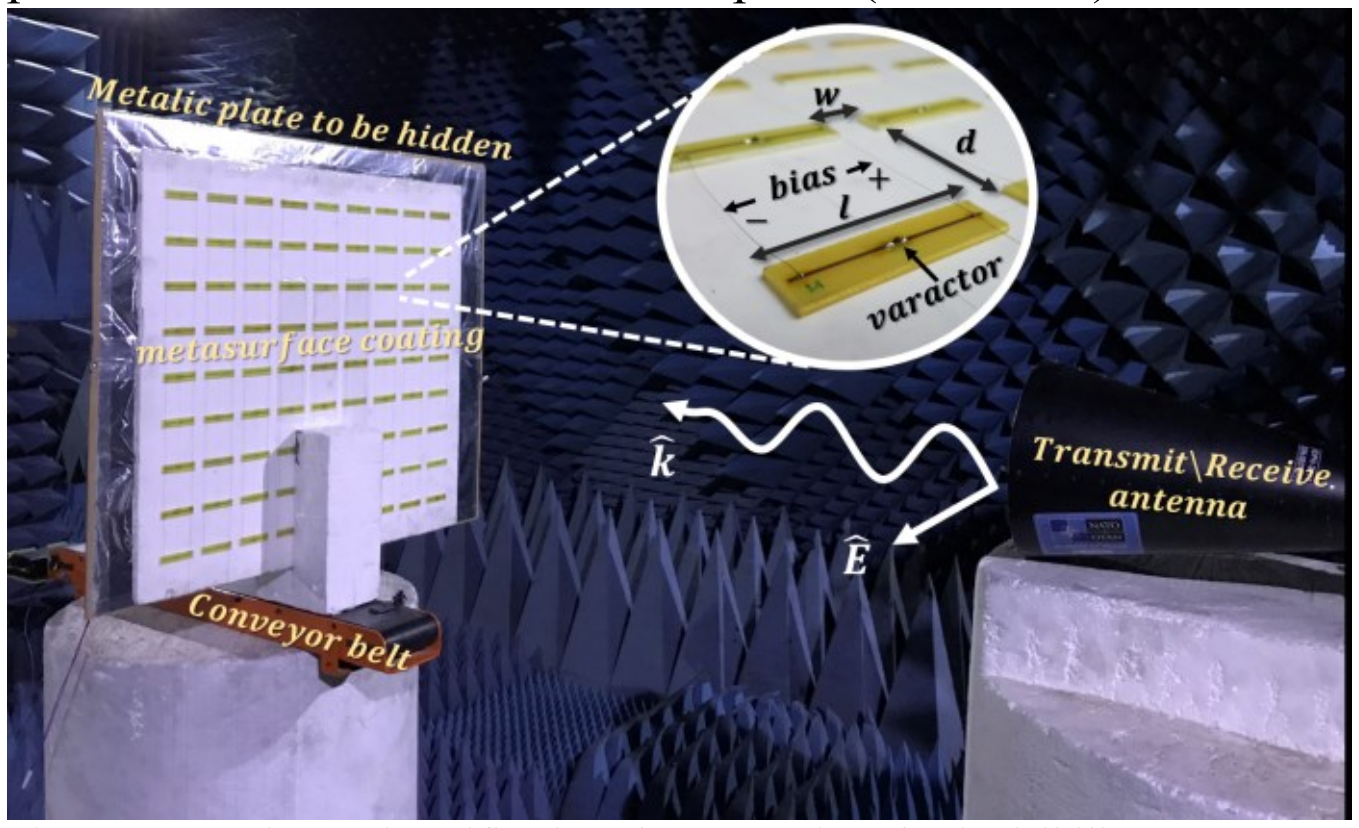

Fig. 4. Experimental verification demonstrating the invisibility concept. A metal plate of 0.25 square meters is covered by a metasurface cloak, which is constructed from an array of dipoles with varactors in their gap, allowing temporally modulating its properties. The concealed target can move on the conveyor belt while being observed by the radar system.

Moving the concealed target without modulating the bias voltage produced linear phase shifts in time as can be seen on Fig. 5(a) and 5(b), which show forward and backward motion correspondingly at constant velocity of 0.03m/s and under 1.5 GHz radar illumination. It can be observed that the full span of $2\pi$ phase was traced in the process of the movement as expected. In the following experiments the target was moved at a constant velocity of -0.03m/s (negative sign indicating motion away from the radar) while various bias voltages were applied to the metasurface. Since the relationship between the modulating voltage and scattered phase is not linear, a calibration procedure was first performed. By applying a linear voltage modulation to the biasing network while keeping the target stationary, the nonlinear temporal phase profile was recorded. Applying the inverse of that function back into the input of the biasing network produced the calibrated linear phase shown on Fig. 5(c), where a maximal phase shift of $\Delta\varphi = 330°$ is achieved, close to the theoretical maximum. It is noted that the phase modulation produced by the metasurface is very similar to the one produced by actual motion, as seen on Fig. 5(b). Controlling the modulation frequency $f_m$ of the metasurface (which is the frequency of the applied periodic voltage drop on the biasing network) while moving the target leads to plots on Fig. 5(d), 5(e), and 5(f). The sign of the frequency indicates the direction of phase modulation (positive and negative signs correspond to increasing and decreasing phase in time, accordingly). The modulation frequency controls the slope of the phase shift $\phi_{Metasurface}(t)$ which is added together with the phase shift produced by the motion of the target via the Doppler Effect as follows:

$$\phi_{tot}(t) = \phi_{Doppler}(t) + \phi_{Metasurface}(t), \quad \phi_{Metasurface}(t) = f_m \Delta\phi t \tag{5}$$

Fig. 5(d), 5(e) and 5(f) show the results for the moving target while different modulation frequencies are applied to the metasurface. In Fig. 5(d) the modulation is of the same polarity as the real motion of the target, leading to their phases adding up as in (5). The result is an overall faster changing slope, corresponding to faster motion than that of the real target. In contrast, Fig. 5(e) has the modulation polarity in the opposite direction to that of the real motion, causing the target to appear slower than it is, as well making it appear to be moving towards the radar while in reality it is moving away from it. The key figure is Fig. 5(f), where the modulation slope is very close to the one created by the real motion of the target, but with opposite polarity, causing almost complete flattening of the phase as a function of time. This corresponds closely to the case $\phi_{Doppler}(t) = -\phi_{Metasurface}(t)$, however slight oscillations of the moving platform in the experiment prevented perfect cancelation. This modulation frequency causes the structure to appear almost stationary to the radar, which means the MTI filter will consider it as clutter, concealing the presence of the target as will be shown ahead. Finally, it is noted that the amplitude of the reflected signal is modulated as well, as seen in Fig. 5(g). Since radar systems tend to rely on the phase of reflected echoes and not their intensity, this amplitude modulation will not affect the results ahead.

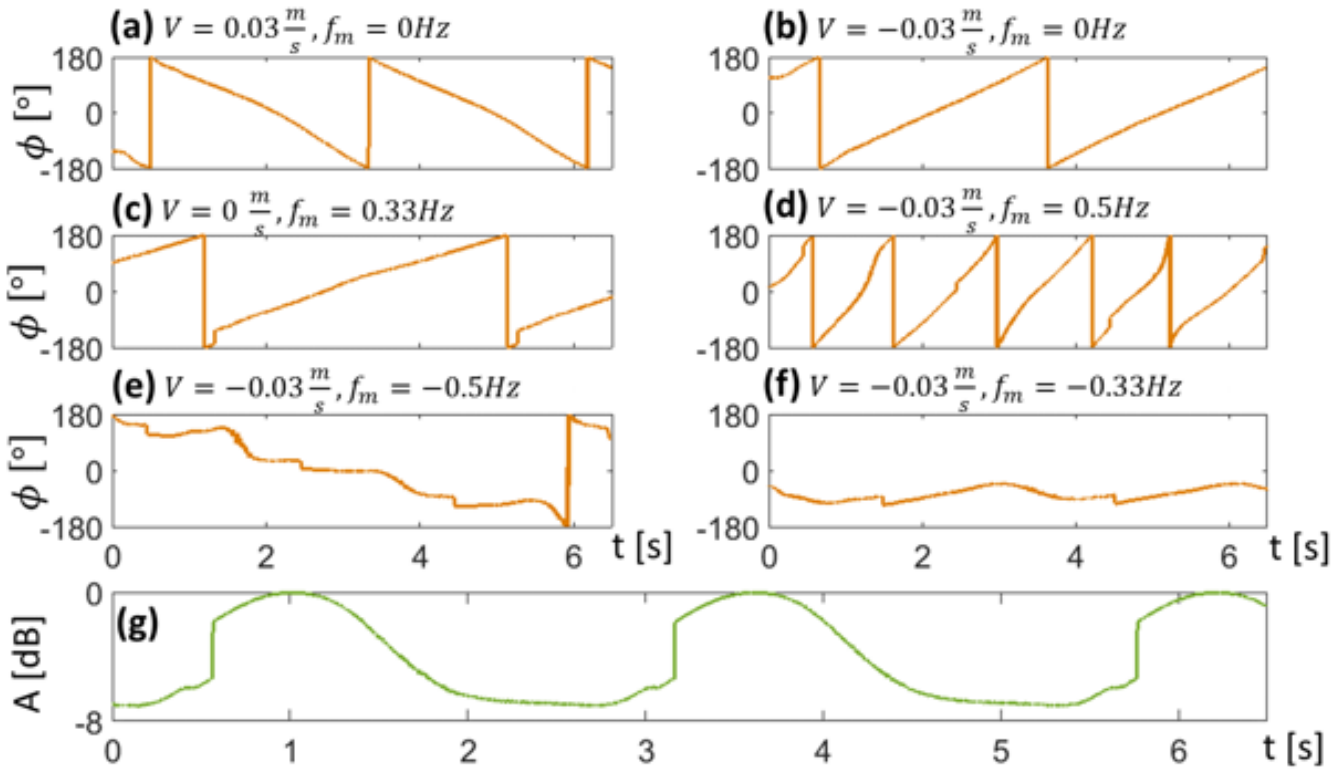

Fig. 5. Analysis of moving targets – phases and amplitudes. $V$ is the true velocity of the structure, $f_m$ is the temporal modulation frequency of the metasurface cover. (a),(b),(c),(d) and (e) – Profiles of the temporal phases retrieved by the radar for various real velocities and metasurface modulation frequencies $f_m$ (negative modulation frequency corresponds to a receding phase slope). (g) The normalized temporal amplitude at the radar receiver for stationary target with modulation $f_m$ = 0.33 Hz.

One of the most popular methods of extracting Doppler information from the phase difference of consecutive pulses is by using a fast Fourier transform (FFT) filter bank. This method, as many other alternatives, serves to average out the signal, which is reminiscent of finding the fittest linear approximation to the function. For real targets in field conditions the phase difference is unlikely to be perfectly linear, partially owing to the fact that the target may fluctuate, e.g. change direction rapidly, enter an area that degrades SNR conditions, or have different moving parts that add additional modulation to the reflected echoes, termed micro-Doppler [37]–[39]. The phases in the plots shown on Fig. 5(a-f) were multiplied by the imaginary unit $j$, exponentiated and put through an FFT filter bank in order to estimate the velocity of the target from the peak of the FFT output, as shown on Fig. 6. Fig. 6(a) shows the output of the FFT for various modulation frequencies with the green dashed line being the output for the moving target without any modulation of the metasurface. The two orange-shade lines correspond to modulation frequencies of $\pm$0.5Hz. For the positive modulation, the output estimates a larger velocity than the ground truth, as shown by the orange dashed outline and in correspondence with the phase in Fig. 5(d). Conversely, for the negative modulation the target appears slower to the radar as well as heading in the opposite direction (towards it), in correspondence with the phase profile of Fig. 5(e). The key result is represented by the red curve on Fig. 6(a), which shows that a modulation frequency of -0.33Hz shifts the perceived velocity of the target to 0, making it appear as stationary to the radar as any of the surrounding clutter. This result is in correspondence with Fig. 5(f). Fig. 6(b) shows the same results as Fig. 6(a) but with the additional processing of an MTI filter implemented as a two-pulse canceller, performed before the FFT processing.

$$y_k = x_k - x_{k-1} \tag{6}$$

where $x_k = e^{i\phi_k}$ and $\phi_k$ are samples of the phase as shown on Fig. 5. Intuitively, the above filter will remove any static contributions of the signal, leaving only time-dependent components that change in between the samples. This can be seen more rigorously by taking the Z-transform of (6). The transfer function of the two pulse canceler MTI is

$$H(z) = 1 - z^{-1} \tag{7}$$

which has zeros at normalized discrete frequencies of $2\pi n$, where $n = 0, 1, 2$ and so on. This means that the DC contribution of the received signal, i.e. the clutter, is removed by the filter. The maximal passband is achieved at normalized discrete frequencies of $\pi(2n+1)$, meaning it is preferable to down-sample the phases in $x_k$ in a way that would allow the expected Doppler frequencies to pass without significant attenuation. This approach was undertaken to produce Fig. 6(b), where the output of the FFT, corresponding to metasurface modulation frequency of -0.33Hz, is completely attenuated. The curves corresponding to modulation frequencies of -0.5Hz and 0Hz are also slightly attenuated by the filter due to their relatively low velocity, while the output of the FFT remains virtually unchanged for the modulation frequency of 0.5Hz. Small side-lobes remain due to the fact the phase is not completely flat, as shown in Fig. 5(f). It is observable that the output of the FFT is somewhat coarse, which is the result of processing only a relatively small time window (and therefore a small amount of samples as per the transfer function restrictions discussed above). This constrain solely relates to the experimental setup, since a short conveyor belt was used and the observation time was limited. It is worth noting that typical settings for airborne target detection are tuned to remove anything moving slowly in the scene, since this can be associated with wind, birds and other nonstationary clutter that will otherwise produce false alarms.

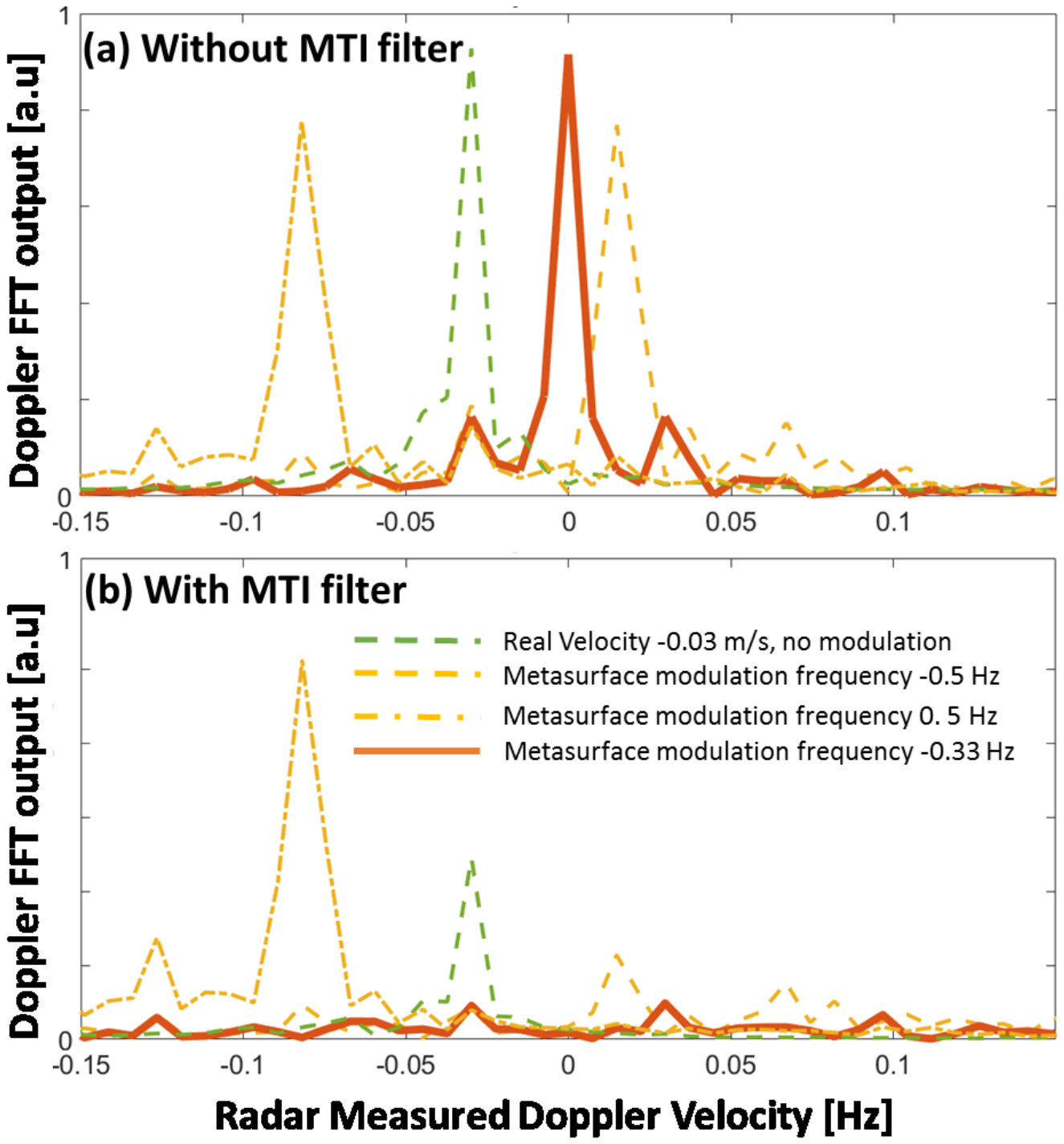

Fig. 6. The output of the radar after Doppler FFT processing (a) without MTI filtering and (b) with MTI filtering. Correct choice of metasurface modulation frequency can shift the measured velocity to 0, which will cause the MTI to filter out the target with the rest of the clutter, making it invisible to the radar.

While Fig. 6 shows the output of the FFT for a select group of metasurface modulation frequencies, it is desirable to demonstrate full control over the velocity measured by the radar. To do so, the experiments mentioned earlier were repeated for 5 different real velocities of the cloaked target, as well as 25 different modulation frequencies, producing Fig. 7. Two important lines are drawn and marked 'real velocity line' and 'invisibility line' respectively. The first indicates the velocity of the target without applying any modulation to the metasurface. The second reveals the required modulation frequency of the metasurface in order to conceal the target (by shifting its effective Doppler signature to 0). A linear relation between the modulation frequency and the measured velocity is observed, in accordance with (5). Some of the data points represented by the green x's on Fig. 7 are the same ones used in Fig. 6, as well as in Fig.5.

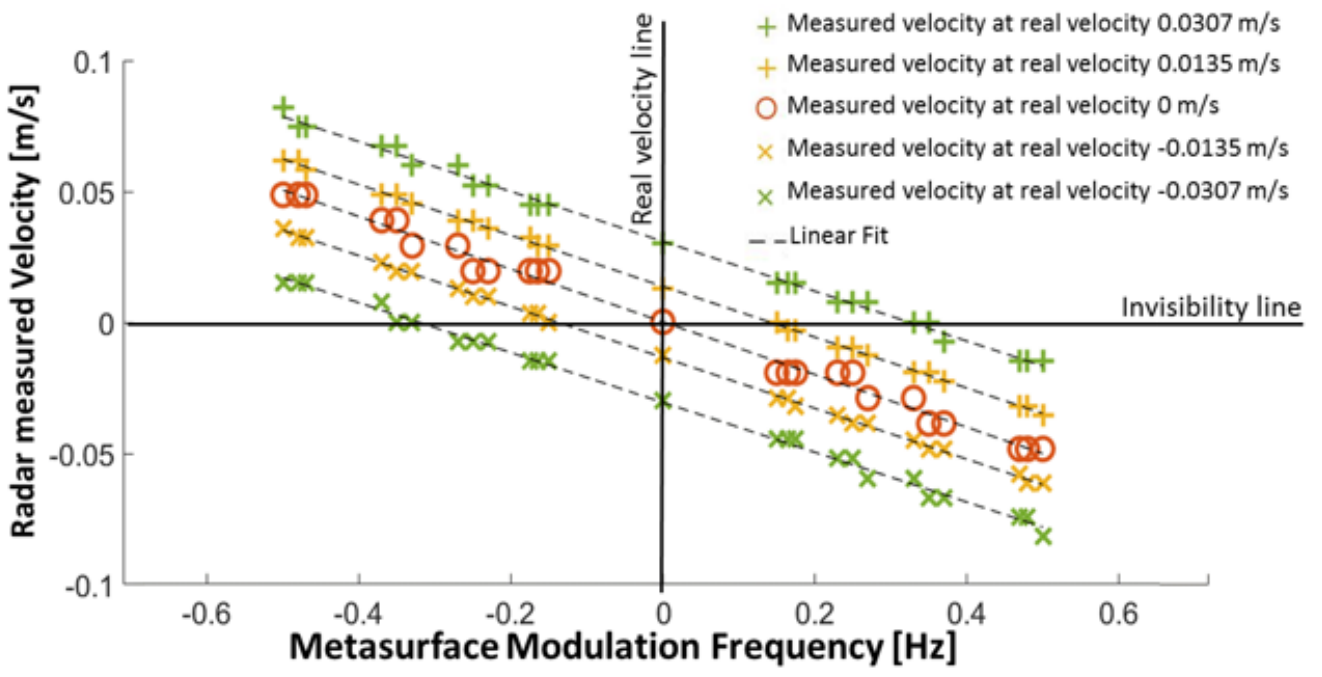

Fig. 7. Measuring velocities from the phase plots by using the FFT method as in (6).

The experiments above were done for a single incident radar carrier frequency of 1.5GHz. The same experiments were repeated for different incident frequencies with similar results, the difference being that a new calibration procedure was required at every frequency due to the fact the threshold capacitance is a function of frequency. If a frequency dependent capacitor is introduced within the design, up to 20% bandwidth can be achieved and a complete concealment of the target will be observed, as it was shown in the simulations on Fig. 3(d).

## IV. Conclusion

The problem of radar invisibility was revisited from a signal processing point of view. While quite a few efforts in the field concentrate on the scattering suppression approach, our proposal demonstrates how to use loopholes in radar's post processing for concealing a target. Specifically, any real radar system uses a filter bank to improve SNR and cope with extremely small echoes from the targets of interest. One of the most dominant sources of noise in the receiver originates from static or slowly moving clutter, which causes substantial backscattered electromagnetic energy. Moving target indicators, regardless of their particular implementation, rely on Doppler information to isolate an object of interest from a clutter. Our approach uses this extremely powerful technique as a weakness. Temporally modulated metasurface covers were shown to be capable of imprinting arbitrary Doppler shifts on to the backscattered echoes. More importantly, they were shown to be able to compensate for real Doppler shifts caused by genuine motion of the target, causing a moving target to look like a stationary one. It worth stressing that any unclassified radar system, operating under real outdoor conditions (not in an isolated environment, e.g. an anechoic chamber) will filter out zero-Doppler targets, even though they reflect quite a substantial amount of energy. This phase-based realization has significant advantages over amplitude approaches, which aim on suppression of reflected energy. For example, reducing the reflection by a factor of 2 will lead to only 3dB SNR reduction, which is quite negligible for most radar systems, operating with 90dB and even higher dynamic ranges. Only dramatic reduction of target's reflection coefficient for a broadband, mixed polarization, all angle of incidence causes successful concealment of the target. Meanwhile, our phase approach already demonstrates perfect cloaking of macroscopic objects.

## Acknowledgment

## References

[1] W. F. Bahret, "The Beginnings of Stealth Tech nology," *IEEE Trans. Aerosp. Electron. Syst.*, vol. 29, no. 4, pp. 1377–1385, 1970.

[2] P. M. Grant and J. H. Collins, "Introduction to electronic warfare," *IEE Proc. F Commun. Radar Signal Process.*, vol. 129, no. 3, pp. 113–132, 1982, doi: 10.1049/ip-f-1.1982.0020.

[3] H. J. Friedman, "Jamming Susceptibility," *IEEE Trans. Aerosp. Electron. Syst.*, vol. AES-4, no. 4, pp. 515–528, 1968, doi: 10.1109/TAES.1968.5409019.

[4] J. V. Difranco and C. Kaiteris, "Radar Performance Review in Clear and Jamming Environments," *IEEE Trans. Aerosp. Electron. Syst.*, vol. AES-17, no. 5, pp. 701–710, 1981, doi:

10.1109/TAES.1981.309102.
[5] F. Capolino, *Applications of Metamaterials*. CRC Press, 2017.
[6] N. I. Zheludev and Y. S. Kivshar, "From metamaterials to metadevices," *Nature Materials*, vol. 11, no. 11. Nature Publishing Group, pp. 917–924, 23-Oct-2012, doi: 10.1038/nmat3431.
[7] N. Engheta and R. Ziolkowski, *Electromagnetic Metamaterials: Physics and Engineering Explorations*. 2006.
[8] *Tutorials in Metamaterials*. CRC Press, 2016.
[9] D. R. Smith, J. B. Pendry, and M. C. K. Wiltshire, "Metamaterials and negative refractive index," *Science*, vol. 305, no. 5685. American Association for the Advancement of Science, pp. 788–792, 06-Aug-2004, doi: 10.1126/science.1096796.
[10] E. Wolf *et al.*, "Controlling Electromagnetic Fields," *Science (80-. ).*, vol. 312, no. June, pp. 1780–1782, 2006.
[11] B. A. Munk, *Frequency Selective Surfaces*. 2000.
[12] D. Schurig *et al.*, "Metamaterial electromagnetic cloak at microwave frequencies," *Science (80-. ).*, vol. 314, no. 5801, pp. 977–980, Nov. 2006, doi: 10.1126/science.1133628.
[13] B. Choudhury and R. M. Jha, "A Review of Metamaterial Invisibility Cloaks."
[14] P. Alitalo and S. Tretyakov, "Electromagnetic cloaking with metamaterials," *Materials Today*, vol. 12, no. 3. Elsevier, pp. 22–29, 01-Mar-2009, doi: 10.1016/S1369-7021(09)70072-0.
[15] R. Fleury and A. Alu, "Cloaking and Invisibility : A Review," *Prog. Electromagn. Res.*, vol. 147, no. February, pp. 171–202, 2014.
[16] A. Alu and N. Engheta, "Achieving transparency with plasmonic and metamaterial coatings," pp. 1–9, 2005, doi: 10.1103/PhysRevE.72.016623.
[17] U. Leonhardt, "Optical conformal mapping," *Science (80-. ).*, vol. 312, no. 5781, pp. 1777–1780, Jun. 2006, doi: 10.1126/science.1126493.
[18] A. Chen and F. Monticone, "On broadband active cloaking," in *2019 IEEE International Symposium on Antennas and Propagation and USNC-URSI Radio Science Meeting, APSURSI 2019 - Proceedings*, 2019, pp. 1317–1318, doi: 10.1109/APUSNCURSINRSM.2019.8888566.
[19] D. Ramaccia, D. L. Sounas, A. Alù, A. Toscano, and F. Bilotti, "Doppler cloak restores invisibility to objects in relativistic motion," *Phys. Rev. B*, vol. 95, no. 7, p. 075113, Feb. 2017, doi: 10.1103/PhysRevB.95.075113.
[20] D. Ramaccia, D. L. Sounas, A. Alu, A. Toscano, and F. Bilotti, "Phase-Induced Frequency Conversion and Doppler Effect with Time-Modulated Metasurfaces," *IEEE Trans. Antennas Propag.*, vol. 68, no. 3, pp. 1607–1617, Mar. 2020, doi: 10.1109/TAP.2019.2952469.
[21] B. Liu, H. Giddens, Y. Li, Y. He, S.-W. Wong, and Y. Hao, "Design and experimental demonstration of Doppler cloak from spatiotemporally modulated metamaterials based on rotational Doppler effect," *Opt. Express*, vol. 28, no. 3, p. 3745, Feb. 2020, doi: 10.1364/oe.382700.
[22] N. Levanon and E. Mozeson, *Radar Signals*. John Wiley & Sons, Inc., 2004.
[23] M. A. Richards, J. Scheer, and W. Holm, *Principles of Modern Radar*. 2010.
[24] R. Klemm, H. Griffiths, and W. Koch, *Novel Radar Techniques and Applications*, vol. 2. 2017.
[25] K. Kvernsveen, "An Adaptive MTI Filter for Coherent Radar," in *IEEE Digital Signal Processing Workshop*, 1996, pp. 351–353.
[26] G. Cloutier, D. Chen, L. Durand, and S. Member, "A New Clutter Rejection Algorithm for Doppler Ultrasound," vol. 22, no. 4, pp. 530–538, 2003.
[27] B. Dawidowicz, K. S. Kulpa, M. Malanowski, J. Misiurewicz, P. Samczynski, and M. Smolarczyk, "DPCA detection of moving targets in airborne passive radar," *IEEE Trans. Aerosp. Electron. Syst.*, vol. 48, no. 2, pp. 1347–1357, Apr. 2012, doi: 10.1109/TAES.2012.6178066.
[28] E. Hyun and J. Lee, "Two – step Moving Target Detection Algorithm for Automotive 77 GHz FMCW Radar," 2010.
[29] V. Kozlov, D. Filonov, A. S. Shalin, B. Z. Steinberg, and P. Ginzburg, "Asymmetric backscattering from the hybrid magneto-electric meta particle," *Appl. Phys. Lett.*, vol. 109, no. 20, p. 203503, Nov. 2016, doi: 10.1063/1.4967238.
[30] V. Kozlov, S. Kosulnikov, D. Filonov, A. Schmidt, and P. Ginzburg, "Coupled micro-Doppler signatures of closely located targets," *Phys. Rev. B*, vol. 100, no. 21, p. 214308, Dec. 2019, doi: 10.1103/PhysRevB.100.214308.
[31] D. H. Werner, *Broadband Metamaterials in Electromagnetics: Technology and Applications*. 2017.
[32] L. Novotny and B. Hecht, *Principles of nano-optics*, vol. 9780521832243. Cambridge University Press, 2006.
[33] Y. Ra'Di, V. S. Asadchy, and S. A. Tretyakov, "Total absorption of electromagnetic waves in ultimately thin layers," *IEEE Trans. Antennas Propag.*, vol. 61, no. 9, pp. 4606–4614, 2013, doi: 10.1109/TAP.2013.2271892.
[34] B. Chambers and A. Tennant, "The phase-switched screen," *IEEE Antennas Propag. Mag.*, vol. 46, no. 6, pp. 23–37, Dec. 2004, doi: 10.1109/MAP.2004.1396733.
[35] J. Long, "Non-Foster Circuit Loaded Periodic Structures for Broadband Fast and Slow Wave Propagation," University of California, San Diego, 2015.
[36] H. Wang *et al.*, "An Elegant Solution: An Alternative Ultra-Wideband Transceiver Based on Stepped-Frequency Continuous-Wave Operation and Compressive Sensing," *IEEE Microw. Mag.*, vol. 17, no. 7, pp. 53–63, Jul. 2016, doi: 10.1109/MMM.2016.2549138.
[37] V. C. Chen, D. Tahmoush, and W. J. Miceli, *Radar Micro-Doppler Signature Processing and applications*. 2014.
[38] H. W. Victor C. Chen, "Micro-doppler effect in radar: phenomenon, model, and simulation study," *IEEE Transactions on Aerospace and Electronic Systems*. .
[39] V. Kozlov, D. Filonov, Y. Yankelevich, and P. Ginzburg, "Micro-Doppler frequency comb generation by rotating wire scatterers," *J. Quant. Spectrosc. Radiat. Transf.*, vol. 190, pp. 7–12, Mar. 2017, doi: 10.1016/j.jqsrt.2016.12.029.